\documentclass[10pt,conference]{IEEEtran}
\IEEEoverridecommandlockouts
\usepackage{cite}
\usepackage{amsmath,amssymb,amsfonts}
\usepackage{algorithmic}
\usepackage{graphicx}
\usepackage{xcolor}

\usepackage[table]{xcolor}

\usepackage{xurl}
\usepackage{subcaption}
\usepackage{multirow}
\usepackage{booktabs}

\usepackage[hidelinks]{hyperref}
\usepackage{stix}

\usepackage{tcolorbox}
\newtcolorbox{boxH}{
    colback = white!90!gray, 
    colframe = black, 
    boxrule = 0pt, 
    leftrule = 3pt
}

\newcommand{\pct}[2]{{\setlength{\fboxsep}{1.5pt}\colorbox{#1}{#2}}}

\usepackage{listings}
\definecolor{codegreen}{rgb}{0,0.6,0}
\definecolor{codegray}{rgb}{0.5,0.5,0.5}
\definecolor{codepurple}{rgb}{0.58,0,0.82}
\definecolor{backcolour}{rgb}{0.95,0.95,0.92}
\lstdefinestyle{mystyle}{
    commentstyle=\color{codegreen},
    keywordstyle=\color{magenta},
    numberstyle=\tiny\color{codegray},
    stringstyle=\color{codepurple},
    basicstyle=\ttfamily\footnotesize,
    breakatwhitespace=false,         
    breaklines=true,                 
    captionpos=b,                    
    keepspaces=true,                 
    numbers=left,                    
    numbersep=5pt,                  
    showspaces=false,                
    showstringspaces=false,
    showtabs=false,                  
    tabsize=2
}

\def\BibTeX{{\rm B\kern-.05em{\sc i\kern-.025em b}\kern-.08em
    T\kern-.1667em\lower.7ex\hbox{E}\kern-.125emX}}
\begin{document}

\title{Do Vibe-Coded Projects Contain Tests?\\An Empirical Study}


 \author{
 \IEEEauthorblockN{Andre Hora}
 \IEEEauthorblockA{
\textit{DCC, UFMG}\\
 Belo Horizonte, Brazil \\
 andrehora@dcc.ufmg.br}
 \and
  \IEEEauthorblockN{João Eduardo Montandon}
 \IEEEauthorblockA{
\textit{DCC, UFMG}\\
 Belo Horizonte, Brazil \\
joao@dcc.ufmg.br}
 \and
  \IEEEauthorblockN{Fabio Petrillo}
 \IEEEauthorblockA{
\textit{École de Technologie Supérieure}\\
 Montreal, Canada \\
fabio.petrillo@etsmtl.ca}
}

\maketitle

\begin{abstract}
With the rapid adoption of generative AI, a new practice known as \emph{vibe coding} has emerged: developers prompt AI tools to write code by describing the desired outcome, often without fully understanding the generated code.
This practice may pose risks, amplified by limited code review and, potentially, by the developer's lack of programming knowledge.
Although \emph{software testing} is a fundamental safeguard, it is unclear whether vibe-coded projects include tests at all.
We present the first empirical study of tests in vibe-coded projects, analyzing 335 GitHub repositories self-declared as ``vibe-coded''.
We find that about half of the vibe-coded projects (52.8\%) include tests, compared with around 90\% in popular GitHub projects.
When present, tests are introduced early, often in the first commit.
Also, test adoption is increasing, rising from 42.9\% in 2025 to 57.4\% in 2026 and reaching 72.2\% in late 2026.
We conclude by discussing implications for research and practice.
\end{abstract}

\begin{IEEEkeywords}
Software testing, Test comprehension, Vibe coding, Generative AI
\end{IEEEkeywords}

\section{Introduction}

Generative AI (GenAI) is transforming how software is designed and developed~\cite{fan2023large, hou2023large}. Coding agents such as Claude Code, Codex, and GitHub Copilot operate with a high degree of \emph{autonomy}: they navigate codebases, edit files, invoke tools, execute code, and complete development tasks with limited human intervention~\cite{agentminingpaper}.
Their adoption has been rapid: recent large-scale studies estimate that by early 2026, coding agents had been adopted in nearly 30\% of GitHub projects overall~\cite{robbes2026agentic} and over 75\% for new projects~\cite{robbes2026verymuchagentic}.

In this context of rapid adoption of AI code generation tools, a new practice known as \emph{vibe coding} has emerged. Coined by Karpathy in 2025,\footnote{\url{https://x.com/karpathy/status/1886192184808149383}} vibe coding is a programming approach in which users prompt AI tools to write code by describing the desired outcome in natural language, without necessarily fully understanding the generated code~\cite{fawzy2026vibe}.
Specifically, vibe coding can be defined as the practice of developing software with AI code generation tools by describing goals in natural language and refining the output through iterative prompting, with little or no review of the generated code~\cite{fawzy2026vibe}.
Vibe coding is practiced both by professional developers, who may use it for rapid prototyping, and by non-developers, who may use it to build applications without programming knowledge~\cite{fawzy2026vibe}. 

A growing body of research has sought to characterize the practice of vibe coding, including its motivations and challenges~\cite{ray2025review, fawzy2026vibe, sapkota2025vibe, sarkar2025vibe, geng2026exploring, huq2026vibecodingproduceaccessible}.
For example, Fawzy et al.~\cite{fawzy2026vibe} found that speed and efficiency are the primary motivations for vibe coding.
Ray~\cite{ray2025review} identified key challenges of vibe coding, including model hallucinations, technical debt, and security risks.
Sarkar and Drosos~\cite{sarkar2025vibe} found that vibe coding follows iterative goal-satisfaction cycles, with prompts that combine vague, high-level directives and detailed technical specifications.

In practice, vibe coding poses several risks, such as technical debt and security vulnerabilities, which are amplified by the limited review of generated code and, in some cases, the developer's lack of programming knowledge~\cite{fawzy2026vibe, ray2025review, huq2026vibecodingproduceaccessible}. Software testing is a fundamental safeguard against such risks, helping prevent defects and sustain software evolution, yet it is unclear whether a practice built on minimal code review leaves any room for it. This raises an open question: \textbf{to what extent are vibe-coded projects tested?}

This short paper presents the first empirical study of tests in vibe-coded projects. We analyze 335 GitHub repositories self-declared as ``vibe-coded'' and examine how often they include tests, when tests are introduced, and how test adoption has evolved.
Specifically, we address three research questions:

\begin{itemize}

    \item \textbf{RQ1: How common are tests in vibe-coded repositories?}
    About half of the vibe-coded repositories (52.8\%, 177 of 335) include tests, compared with around 90\% of popular projects.
    Moreover, only 43.6\% of the vibe-coded repositories have CI/CD files, and 70\% of these mention tests; in popular projects, both rates are around 90\%.
    
    \item \textbf{RQ2: When are tests introduced in vibe-coded repositories?}
    Tests are introduced early: 48.6\% have tests from the first commit.
    Moreover, in vibe-coded projects with tests, one in four commits changes test files. 
    
    \item \textbf{RQ3: How has the adoption of tests in vibe-coded repositories evolved over time?}
    Test adoption in vibe-coded projects is increasing over time: the proportion of projects with tests rose from 42.9\% in 2025 to 57.4\% in 2026, reaching 72.2\% in late 2026.

\end{itemize}

\noindent\textbf{Contributions.}
The contributions of this paper are threefold.
First, we present the first empirical study of tests in vibe-coded projects.
Second, we discuss the implications of our findings for research and practice.
Third, we make our dataset of vibe-coded repositories publicly available.

\noindent\textbf{Dataset.}
Our dataset is publicly available at: \url{https://doi.org/10.5281/zenodo.24568931}

\section{Study Design}

\subsection{Detecting Vibe-Coded Repositories}
\label{sec:detection}

Our goal is to analyze vibe-coded projects; however, reliably identifying them is not trivial for several reasons.
First, developers may not disclose the use of AI code generation tools in commit messages~\cite{agentminingpaper}.
Second, developers may use such tools but commit the generated code under their own accounts~\cite{agentminingpaper}.
Third, even when AI-generated code can be detected, it remains unclear whether the project was vibe-coded or developed with AI assistance following conventional engineering practices~\cite{sapkota2025vibe}.
To overcome these challenges, we focus on projects self-declared as ``vibe-coded'', that is, projects whose creators explicitly describe them as vibe-coded.

We used the GitHub Search API to collect repositories that self-identify as vibe-coded. A repository was a candidate if it had the topic \emph{vibe-coded} or \emph{vibecoded}, or if its description contained the term \emph{vibe coded}, \emph{vibe-coded}, or \emph{vibecoded}.
We considered only repositories created on or after February, 2025, the date the term was coined.
To exclude abandoned or trivial projects, we required at least 5 stars and left out archived repositories.
Finally, we inspected the matches and excluded repositories that mention the terms but are not vibe-coded software projects of their own, such as guides, collections, and other non-code artifacts.
After this filtering step, we obtained a final dataset of 335 vibe-coded repositories.

\subsection{Characterization of the Vibe-Coded Repositories}

Our dataset comprises 335 vibe-coded repositories.
The median number of stars is 9 (Q1 = 6, Q3 = 20), ranging from 5 to 1,897. 
Despite being vibe-coded, some of these projects have become popular: 21 repositories have more than 100 stars, and two exceed 1,000.
The median number of commits is 35 (Q1 = 8, Q3 = 134).
The repositories are written in 27 primary languages, led by Python and TypeScript (60 repositories each), followed by JavaScript (46) and Rust (45).


\subsection{RQ1: Frequency of Tests in Vibe-Coded Repositories}

This RQ provides insights into the presence of tests in vibe-coded projects, focusing on test files and on tests in CI/CD.

\smallskip

\noindent\textbf{Test Files.}
We identify test files from their paths.
A file counts as a test file when it has a source-code extension (any programming language in GitHub Linguist) and meets at least one of two conditions.
(1) Its name follows a common test naming convention, such as a test prefix or suffix or a .test/.spec infix.
(2) It is located in a test directory, such as \texttt{test/}, \texttt{tests/}, \texttt{\_\_tests\_\_/}, or \texttt{spec/}.

\smallskip

\noindent\textbf{Tests in CI/CD.}
We also analyze the presence of five widely used CI/CD tools~\cite{decan2022use}: GitHub Actions, Travis CI, CircleCI, Azure Pipelines, and Jenkins.
A repository uses CI/CD when it contains a configuration file for at least one of them.
We consider that its CI/CD mentions tests when a configuration file is named after tests or when a non-comment line contains the word test, tests, or testing.

Finally, for comparison, we report the corresponding rates for four groups of real-world software repositories (control, recently popular, popular, and very popular), each containing 335 repositories.
\emph{Control:} a control group, with repositories randomly selected from GitHub with similar language, commit, creation date, and star distributions to those of the vibe-coded ones.
\emph{Recently popular:} the most-starred repositories on GitHub with similar language and creation date distributions to those of the vibe-coded ones.
\emph{Popular:} the most-starred repositories on GitHub with a similar language distribution to that of the vibe-coded ones, regardless of creation date.
\emph{Very popular:} the most-starred repositories on GitHub.

\subsection{RQ2: Test Activity in Vibe-Coded Repositories}
In this RQ we explore the test activity in the vibe-coded projects.
Specifically, for each vibe-coded repository, we analyze its commit history from the oldest to the newest commit. A test commit is a commit that adds, modifies, or deletes a test file. We compute the share of test commits, the position of the first test commit in the history, and the number of days from the first commit to the first test commit.

\subsection{RQ3: Evolution of Tests in Vibe-Coded Repositories}

Finally, we analyze the presence of tests over time. We group the vibe-coded repositories by their \emph{creation date} into calendar years, six-month periods, and two-month periods. For each period, we compute the proportion of repositories that include at least one test file.
Our goal is to understand whether testing practices in vibe-coded repositories have changed as AI coding tools have evolved and become more widely adopted.

\section{Results}

\subsection{RQ1: Frequency of Tests in Vibe-Coded Repositories}

\noindent\textbf{Test Files.}
Overall, we find that 52.8\% (177 of 335) of the vibe-coded repositories include tests.
For comparison, the Table~\ref{tab:has-tests} reports the corresponding rates for a control group and for three groups of real-world repositories (recently popular, popular, and very popular).

Vibe-coded repositories include tests at a rate similar to that of the control group (52.8\% vs.\ 57.0\%).\footnote{Fisher's exact test vs. vibe-coded (Holm-adjusted p-values); effect size: phi with Cohen's thresholds.}
However, they are significantly less likely to include tests than popular repositories, whose rates are 88.4\% for recently popular, 91.6\% for popular, and 94.6\% for very popular repositories.

\begin{table}[h]
\centering
\caption{Repositories with test files. P-values and effect sizes are relative to the vibe-coded repositories.}
\label{tab:has-tests}
\small
\begin{tabular}{l r r r l}
\toprule
\textbf{Group} & \textbf{Repos} & \textbf{With tests} & \textbf{p-value} & \textbf{Effect} \\
\midrule
Vibe-coded   & 177/335 & \pct{yellow!31!red!40}{52.8\%}   & --       & --         \\
\midrule
Control      & 191/335 & \pct{yellow!48!red!40}{57.0\%}   & 0.313    & -- \\
Recently popular       & 296/335 & \pct{green!74!yellow!40}{88.4\%} & $<$0.001 & medium     \\
Popular      & 307/335 & \pct{green!86!yellow!40}{91.6\%} & $<$0.001 & medium     \\
Very popular & 317/335 & \pct{green!98!yellow!40}{94.6\%} & $<$0.001 & medium     \\
\bottomrule
\end{tabular}
\end{table}

Table~\ref{tab:test-files} complements this analysis with the median number of test files, considering only repositories that include tests.
Vibe-coded projects have a median of 13 test files, similar to the control group (15). Popular projects have considerably more test files, with medians ranging from 234.5 to 327.

\begin{table}[h]
\centering
\caption{Number of test files in repositories with tests (Q1: first quartile; Q3: third quartile).}
\label{tab:test-files}
\small
\begin{tabular}{l r r r r}
\toprule
 & & \multicolumn{3}{c}{\textbf{Test files}} \\
\cmidrule(lr){3-5}
\textbf{Group} & \textbf{Repos} & \textbf{Q1} & \textbf{Median} & \textbf{Q3} \\
\midrule
Vibe-coded   & 177 & 3     & 13    & 61     \\
\midrule
Control      & 191 & 3     & 15    & 48.5   \\
Recently popular       & 296 & 48.75 & 234.5 & 691    \\
Popular      & 307 & 56.5  & 255   & 1438.5 \\
Very popular & 317 & 62    & 327   & 1597   \\
\bottomrule
\end{tabular}
\end{table}

We also analyze the vibe-coded repositories by programming language, as shown in Table~\ref{tab:has-tests-language}. Although 52.8\% of vibe-coded repositories include tests overall, this proportion varies widely by language, ranging from 20.8\% (HTML) to 76.9\% (Go). Even among the most common languages, the rates differ considerably: Python and Rust exceed the overall rate (61.7\% and 68.9\%, respectively), whereas TypeScript and JavaScript fall well below it (40.0\% each).

\begin{table}[h]
\centering
\caption{Vibe-coded repositories by language.}
\label{tab:has-tests-language}
\small
\begin{tabular}{l r r}
\toprule
\textbf{Language} & \textbf{Repos} & \textbf{With tests} \\
\midrule
Python                & 37/60   & 61.7\% \\
TypeScript            & 24/60   & 40.0\% \\
Rust                  & 31/45   & 68.9\% \\
JavaScript            & 18/45   & 40.0\% \\
HTML                  & 5/24    & 20.8\% \\
C++                   & 9/16    & 56.2\% \\
C                     & 9/15    & 60.0\% \\
Go                    & 10/13   & 76.9\% \\
Kotlin                & 8/12    & 66.7\% \\
Swift                 & 6/9     & 66.7\% \\
Others (17 languages) & 20/36   & 55.6\% \\
\midrule
Total                 & 177/335 & 52.8\% \\
\bottomrule
\end{tabular}
\end{table}

\begin{boxH}
\textbf{Finding 1:} About half of the vibe-coded repositories (52.8\%, 177 of 335) include tests, compared with around 90\% of popular projects.
\end{boxH}

\smallskip

\noindent\textbf{Tests in CI/CD.}
Table~\ref{tab:cicd} details the presence of CI/CD files in the analyzed repositories. Less than half (43.6\%) of the vibe-coded repositories contain CI/CD files, compared with 91.9\% of very popular projects. Moreover, among repositories with CI/CD files, only 69.2\% of vibe-coded projects reference tests in their CI/CD configuration, compared with 92.2\% of very popular projects.

\begin{table}[h]
\centering
\caption{Repositories with CI/CD files.}
\label{tab:cicd}
\small
\begin{tabular}{l r r r r}
\toprule
 & \multicolumn{2}{c}{\textbf{Has CI/CD files}} & \multicolumn{2}{c}{\textbf{Mention test in CI/CD}} \\
\cmidrule(lr){2-3} \cmidrule(lr){4-5}
\textbf{Group} & \textbf{Repos} & \textbf{\%} & \textbf{Repos} & \textbf{\%} \\
\midrule
Vibe-coded   & 146/335 & \pct{yellow!0!red!40}{43.6\%}    & 101/146 & \pct{yellow!97!red!40}{69.2\%}   \\
\midrule
Control      & 176/335 & \pct{yellow!30!red!40}{52.5\%}   & 136/176 & \pct{green!29!yellow!40}{77.3\%} \\
Recently popular       & 292/335 & \pct{green!69!yellow!40}{87.2\%} & 254/292 & \pct{green!68!yellow!40}{87.0\%} \\
Popular      & 301/335 & \pct{green!80!yellow!40}{89.9\%} & 272/301 & \pct{green!82!yellow!40}{90.4\%} \\
Very popular & 308/335 & \pct{green!88!yellow!40}{91.9\%} & 284/308 & \pct{green!89!yellow!40}{92.2\%} \\
\bottomrule
\end{tabular}
\end{table}

\begin{boxH}
\textbf{Finding 2:} Only 43.6\% of the vibe-coded repositories have CI/CD files, and 70\% of these mention tests; in popular projects, both rates are around 90\%.
\end{boxH}

\subsection{RQ2: Test Activity in Vibe-Coded Repositories}

Table~\ref{tab:test-commits} presents the test activity in the vibe-coded repositories that include test files. These repositories have a median of 79 commits, of which 11 change test files. The median proportion of commits that change test files is 25\%; that is, in a typical repository, about one in four commits touches the tests.
Once present, tests are not simply added once and left untouched, but are updated regularly as the code evolves.

Tests are also introduced early in the development history. At the median, the first test appears in the 2nd commit, on the same day as the first commit.
Nearly half of these repositories (48.6\%, 86 of 177) include tests in their very first commit.

\begin{table}[h]
\centering
\caption{Test activity in vibe-coded repositories with tests.}
\label{tab:test-commits}
\small
\begin{tabular}{l r}
\toprule
\textbf{Metric} & \textbf{Vibe-coded} \\
\midrule
Commits (median)                                   & 79     \\
Commits changing test files (median)            & 11     \\
Ratio of commits changing test files (median)          & 25.0   \\ \midrule
Commit number of the first test (median)           & 2      \\
Days from first commit to first test (median)      & 0      \\
Repos with tests in the first commit      & 86/177 (48.6\%) \\
\bottomrule
\end{tabular}
\end{table}

\begin{boxH}
\textbf{Finding 3:} In vibe-coded projects with tests, one in four commits changes test files. Tests are also introduced early: 48.6\% have tests from the first commit.
\end{boxH}

\subsection{RQ3: Evolution of Tests in Vibe-Coded Repositories}

\begin{figure}[h]
\centering
\begin{subfigure}[b]{0.2\textwidth}
  \centering
  \includegraphics[width=\linewidth]{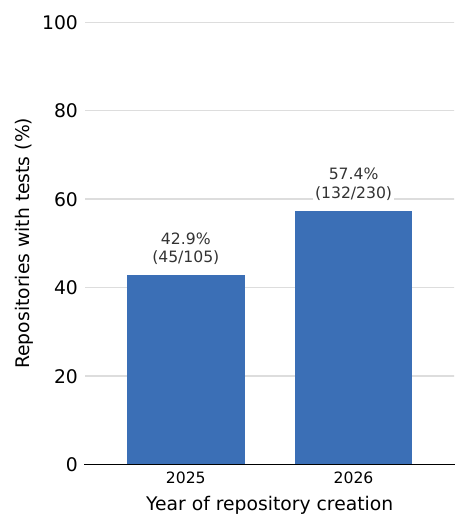}
  \caption{By year}
  \label{fig:tests-by-year}
\end{subfigure}
\hspace{0.5em}
\begin{subfigure}[b]{0.22\textwidth}
  \centering
  \includegraphics[width=\linewidth]{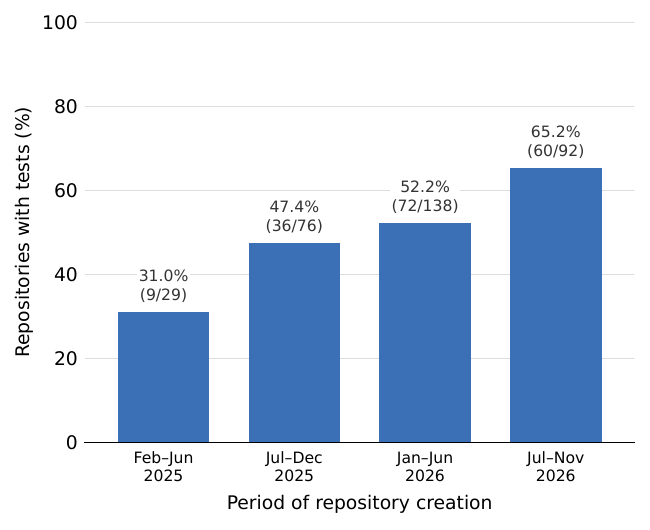}
  \caption{By 6-month period}
  \label{fig:tests-by-6-months}
\end{subfigure}

\vspace{0.5em}

\begin{subfigure}[b]{0.44\textwidth}
  \centering
  \includegraphics[width=\linewidth]{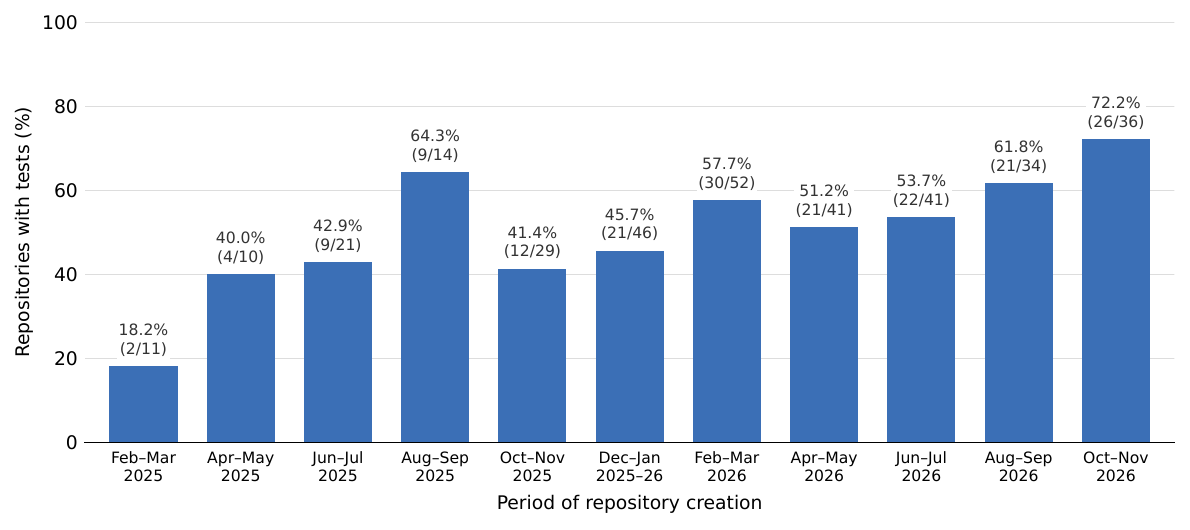}
  \caption{By 2-month period}
  \label{fig:tests-by-2-months}
\end{subfigure}
\caption{Vibe-coded repositories by period of creation.}
\label{fig:tests-over-time}
\end{figure}

Figure~\ref{fig:tests-over-time} shows how the presence of tests in vibe-coded repositories has changed over time. The proportion of repositories with tests increases steadily: from 42.9\% for repositories created in 2025 to 57.4\% for those created in 2026 (Figure~\ref{fig:tests-over-time}a).

The 6-month view confirms this trend, with the proportion rising consistently from 31.0\% (Feb--Jun 2025) to 65.2\% (Jul--Nov 2026) (Figure~\ref{fig:tests-over-time}b).
The 2-month view reveals the overall upward trend holds, rising from 18.2\% in Feb--Mar 2025 to 72.2\% in Oct--Nov 2026, the highest proportion across all periods (Figure~\ref{fig:tests-over-time}c). 

\begin{boxH}
\textbf{Finding 4:} Test adoption in vibe-coded projects is increasing over time: the proportion of projects with tests rose from 42.9\% in 2025 to 57.4\% in 2026.
\end{boxH}

\section{Discussion and Implications}

\subsection{Characteristics of Tests in Vibe-Coded Projects}

Overall, about half of the vibe-coded repositories contain test artifacts (52.8\%, 177 of 335).
Moreover, 43.6\% of them have CI/CD files, and 70\% of these mention tests. On the one hand, these rates can be considered high given the core characteristics of vibe-coded projects: limited review of the generated code and, in some cases, the developer's lack of programming knowledge~\cite{fawzy2026vibe, ray2025review}. On the other hand, the corresponding rates in real-world projects are around 90\%. This shows that, although a substantial share of vibe-coded projects include tests, they still lag behind real-world projects.

\subsection{Vibe-Coded Projects Are Increasingly Likely to Have Tests}

RQ3 reveals an interesting insight: test adoption in vibe-coded projects is increasing over time. The proportion of projects with tests rose from 42.9\% in 2025 to 57.4\% in 2026, reaching 72.2\% in late 2026. This trend may be explained by several factors. \emph{First}, AI coding tools have evolved from code assistants into autonomous agents~\cite{robbes2026agentic, robbes2026verymuchagentic} that often write and run tests~\cite{haque2026empirical, watanabe2026use, zhang2026enhancing, hora2026coding, ardic2026uncovering} to verify their own changes, so tests may increasingly be produced as a by-product of the agent's workflow rather than at the developer's request. 
\emph{Second}, the underlying models have improved, and generating tests alongside code may have become part of their default behavior. 
\emph{Third}, as vibe coding matures, its practitioners may be growing more aware of its risks bringing established testing habits with them.
Investigating which of these factors drives the trend is an interesting direction for future work.

\subsection{Having Tests Does Not Imply Having Good Tests}

This study does not assess the quality of the tests found in vibe-coded projects.
Like production code, test code should follow best practices~\cite{aniche2022effective, yu2019comprehending, aniche2021developers}.
This is an important aspect, since recent studies have shown that test cases generated by LLMs and coding agents may suffer from quality issues~\cite{alshahwan2024automated}, such as test smells~\cite{ouedraogo2024test, alves2024detecting}, over-mocking~\cite{hora2026coding}, and lack of behavioral fidelity~\cite{ning2026executable}.
Tests in vibe-coded projects may have similar problems, given the limited programming experience of their developers.
The presence of tests in vibe-coded projects should not be taken as evidence that they are adequately used.
Understanding what these tests cover and assessing their quality emerge as promising research lines.

\section{Limitations}

We analyze only repositories whose creators explicitly describe them as vibe-coded (Section~\ref{sec:detection}).
We rely on self-declaration because identifying vibe-coded projects is not trivial, for example, developers may not disclose their use of AI tools~\cite{agentminingpaper}.
Our dataset construction therefore favors precision over recall: we likely miss vibe-coded projects, but those we include are explicitly labeled as such by their creators.
Also, we detect tests by file path, so we may miss tests written inside source files (e.g., Rust unit tests).
Lastly, developers who publicly label their projects as vibe-coded may differ from those who do not, so our findings may not generalize to all vibe-coded projects.

\section{Related Work}


Ideally, test suites should be of high quality, so that they protect against regressions and support sustainable software evolution~\cite{aniche2022effective, yu2019comprehending, aniche2021developers, dalton2020exceptional, marcilio2021java, hora2025exceptional}.
Multiple studies investigated the presence of tests in open source projects~\cite{kochhar2013empirical, madeja2021empirical, pan2026hamster}.

Recent studies have investigated tests generated by large language models (LLMs) and coding agents.
Overall, the results are promising; for example, LLM-generated tests can increase coverage~\cite{schafer2023empirical, ryan2024code, chen2024chatunitest}, support TDD~\cite{mathews2024test, ahmed2024tdd}, enhance existing tests~\cite{alshahwan2024automated}, and generate end-to-end tests~\cite{alian2025feature}.
On the other hand, test cases generated by LLMs and coding agents may suffer from quality issues~\cite{ouedraogo2024test, alves2024detecting, alshahwan2024automated, hora2026coding, ning2026executable}.
Recent studies have also explored the presence of test guidelines in agent context files~\cite{ardic2026uncovering, santos2026decoding, hora2026coding, voria2026they}, which may affect the quality of generated tests.
Our study contributes to such research line by exploring the presence of tests in vibe-coded projects. 






\section{Conclusion and Further Work}

We presented the first empirical study of tests in vibe-coded projects, analyzing 335 self-declared vibe-coded GitHub repositories.
When present, tests are introduced early, often in the first commit.
About half of them (52.8\%) included tests, but test adoption was rising, reaching 72.2\% in late 2026. 

As future work, we plan to investigate why test adoption in vibe-coded projects is increasing, for example, due to more capable autonomous agents, improved underlying models, or evolving practices.
We also plan to assess the quality of the generated tests in more depth, such as whether they follow best practices and avoid test smells.


\bibliographystyle{IEEEtran}
\bibliography{main}

\end{document}